\documentclass[aps,prb,superscriptaddress,twocolumn,showkeys,amsmath,amssymb,floatfix]{revtex4-2}

\usepackage{comment}
\usepackage{ulem}
\usepackage[utf8]{inputenc}
\usepackage{graphicx}
\usepackage{dcolumn}
\usepackage{bm}
\usepackage{hyperref}
\usepackage{amsmath}
\usepackage{mathtools}
\usepackage[dvipsnames]{xcolor}
\usepackage{amssymb}
\usepackage{outline}
\definecolor{darkblue}{RGB}{0,0,139} 

\begin{document}

\title{ Finite-size and quenching effects on hyperuniform structures formed during cooling
}

\author{A. Cruz-García}
\affiliation{Low Temperatures Lab, Centro At\'{o}mico Bariloche, CNEA, Argentina.}
\affiliation{Instituto de Nanociencia y Nanotecnología, CONICET-CNEA, Nodo Bariloche,  Argentina.}

\author{J. Puig}
\affiliation{Low Temperatures Lab, Centro At\'{o}mico Bariloche, CNEA, Argentina.}
\affiliation{Instituto de Nanociencia y Nanotecnología, CONICET-CNEA, Nodo Bariloche,  Argentina.}
\affiliation{Instituto Balseiro,
CNEA and Universidad Nacional de Cuyo, 
Bariloche, Argentina.}

\author{R. M. Besana}
\affiliation{Low Temperatures Lab, Centro At\'{o}mico Bariloche, CNEA, Argentina.}
\affiliation{Instituto Balseiro,
CNEA and Universidad Nacional de Cuyo, 
Bariloche, Argentina.}

\author{A. B. Kolton}
\affiliation{Condensed Matter Theory group, Centro At\'{o}mico Bariloche, CNEA,  Argentina}
\affiliation{Instituto Balseiro,
CNEA and Universidad Nacional de Cuyo, 
Bariloche, Argentina.}

\author{Y. Fasano}
\affiliation{Low Temperatures Lab, Centro At\'{o}mico Bariloche, CNEA, Argentina.}
\affiliation{Instituto de Nanociencia y Nanotecnología, CONICET-CNEA, Nodo Bariloche,  Argentina.}
\affiliation{Instituto Balseiro,
CNEA and Universidad Nacional de Cuyo, 
Bariloche, Argentina.}

\date{\today}

\begin{abstract}

The outstanding physical properties of hyperuniform condensed matter systems holds significant promise for technological applications and studying effects that may disrupt this hidden order is therefore very important.  Vortex matter in superconductors is a model system to study this problem since 
imaging experiments have revealed that correlated disorder in the host media and finite size effects disrupt the hyperuniformity of the in-plane arrangement of vortices. Here we report  simulations of layered interacting elastic lines as a model for the vortex lattice in three-dimensional superconductors, following a cooling protocol that closely mimics the experimental conditions. 
We show that finite-thickness effects limiting the hyperuniformity range arise both in equilibrium and out-of-equilibrium. Our results provide a theoretical framework to draw a realistic road-map on synthesizing hyperuniform materials when cooling structures on finite host media with disorder.
\end{abstract}

\maketitle

\section*{Introduction}

Hyperuniformity is a distinctive state of matter characterized by macroscopic homogeneity in the density of constituents of a system~\cite{Gabrielli2003,Torquato2003}.  While such uniformity is intrinsic to crystals, disordered structures can also exhibit a strong suppression of long-wavelength density fluctuations, reflecting hidden long-range correlations~\cite{Torquato2018}. Disordered hyperuniform systems~\cite{Chen2023,Wang2025} have attracted growing attention in fundamental research since their unusual photonic, electronic, thermal, and mechanical properties make them promising candidates for novel technologies.  
Unlike crystals, tailored hyperuniform systems can block light in all directions~\cite{Man2013}. Two-dimensional silica, which is insulating in crystalline form at room temperature, enhance its electronic conduction and become metallic when grown as a disordered hyperuniform structure~\cite{Zheng2020}.  Similarly, disordered hyperuniform carbon monolayers, created by introducing topological defects in graphene, exhibit enhanced thermal conductivity; theoretical studies suggest that the hidden long-range correlations in these systems support extended modes that facilitate heat transport~\cite{Liang2024}. Disordered hyperuniform medium- or high-entropy alloys also display enhanced electronic bandgaps and thermal transport compared with their ordered counterparts~\cite{Chen2023}.  These examples demonstrate that hyperuniform states of matter are not only ubiquitous in nature~\cite{Jiao2014,Dreyfus2015,Chen2018,Torquato2018,Rumi2019,Zheng2020,Llorens2020,Zheng2020,Nizam2021,Chen2021,Chieco2021,Zhang2022,Aragon2023,Philcox2023} but also of considerable interest for the development of cutting-edge technologies. Moreover, the ability to design hyperuniform patterns with targeted properties remains an important challenge in current research~\cite{MilorSalvalaglio2025}.

In hyperuniform systems, the fluctuations in the density of constituents are suppressed at large wavelengths in the asymptotic limit~\cite{Torquato2003}. Consequently, their structure factor typically vanishes as $S(q) \propto q^{\alpha}$ when the reciprocal-space wave vector $q$ approaches $0$, motivating a classification of hyperuniform systems according to ranges of $\alpha$~\cite{Torquato2003}.  According to the fluctuation-compressibility theorem, at equilibrium the value of $S(q=0)$ is proportional to the compressibility of a system~\cite{PathriaBeale2011}. Therefore, theoretically, hyperuniformity can arise at equilibrium only in incompressible systems  at positive finite temperatures ~\cite{Torquato2018}. 
On the other hand, hyperuniformity can also arise out-of-equilibrium, where the fluctuation–compressibility theorem does not hold \cite{Young2023PNAS120,ChenKlattFredrickson2024Macromolecules57}.

In general, incompressibility at thermal  equilibrium can be achieved only in systems with long-range repulsive interactions between constituents~\cite{Torquato2018}.  
Nevertheless, hyperuniform point patterns can exist at equilibrium within higher-dimensional systems with short-range interactions. Indeed, the point pattern formed by the tips of superconducting vortices impinging on the surface of a thick three-dimensional sample with point disorder is an example of this situation~\cite{Rumi2019,Llorens2020,Aragon2023}. If the disorder in the host medium is correlated, the point pattern can become non-hyperuniform with an anisotropic structure factor~\cite{Puig2022,Puig2023} or even anti-hyperuniform~\cite{Llorens2020b,Puig2024}. However, in the case of a host medium with weak uncorrelated disorder, the hidden  hyperuniform correlations arise from an effective long-range interaction within the plane, mediated by the elasticity of vortices across the sample thickness~\cite{Rumi2019}. Interestingly, at equilibrium, due to this effective long-range interaction, the nucleation of hyperuniform vortex patterns is consistent with the fluctuation-compressibility theorem. This has origin in density fluctuations of the vortex tips being associated with the compressibility of a single plane that has a bulk  tilting energy cost.

The bulk-mediated long-range repulsive interactions between vortex tips naturally raise the question of finite-thickness effects. If vortex lines are too short (small thickness $t$) or too rigid, these effective interactions become short-ranged, as the tilt elastic energy contributing to the pair interaction between tips is reduced. This highlights the potential impact of finite-size effects on large-scale structural properties, which is crucial for designing hyperuniform materials.  In this respect, some of us previously used vortex matter in type-II superconductors to study the effects of finite size on hyperuniformity~\cite{Besana2024}. These experiments showed that hyperuniformity degrades when decreasing the thickness of samples with point disorder:  the hyperuniformity exponent $\alpha$ decreases when $t$ decreases if fitted starting from the smallest accessible wave vectors (largest accessible field of view).  Notably, these observations were obtained from snapshots of structures frozen during cooling~\cite{Fasano1999}, performed at the same density and interaction between constituents (magnetic field) but for different thicknesses~\cite{Fasano2003}. This raises the central question motivating this work: Is the thickness dependence of hyperuniformity an equilibrium effect, as predicted in Refs.~\cite{Rumi2019,Besana2024}, or an out-of-equilibrium effect arising from the slow dynamics during cooling?

\begin{figure}[ttt]
\centering
\includegraphics[width=\linewidth]{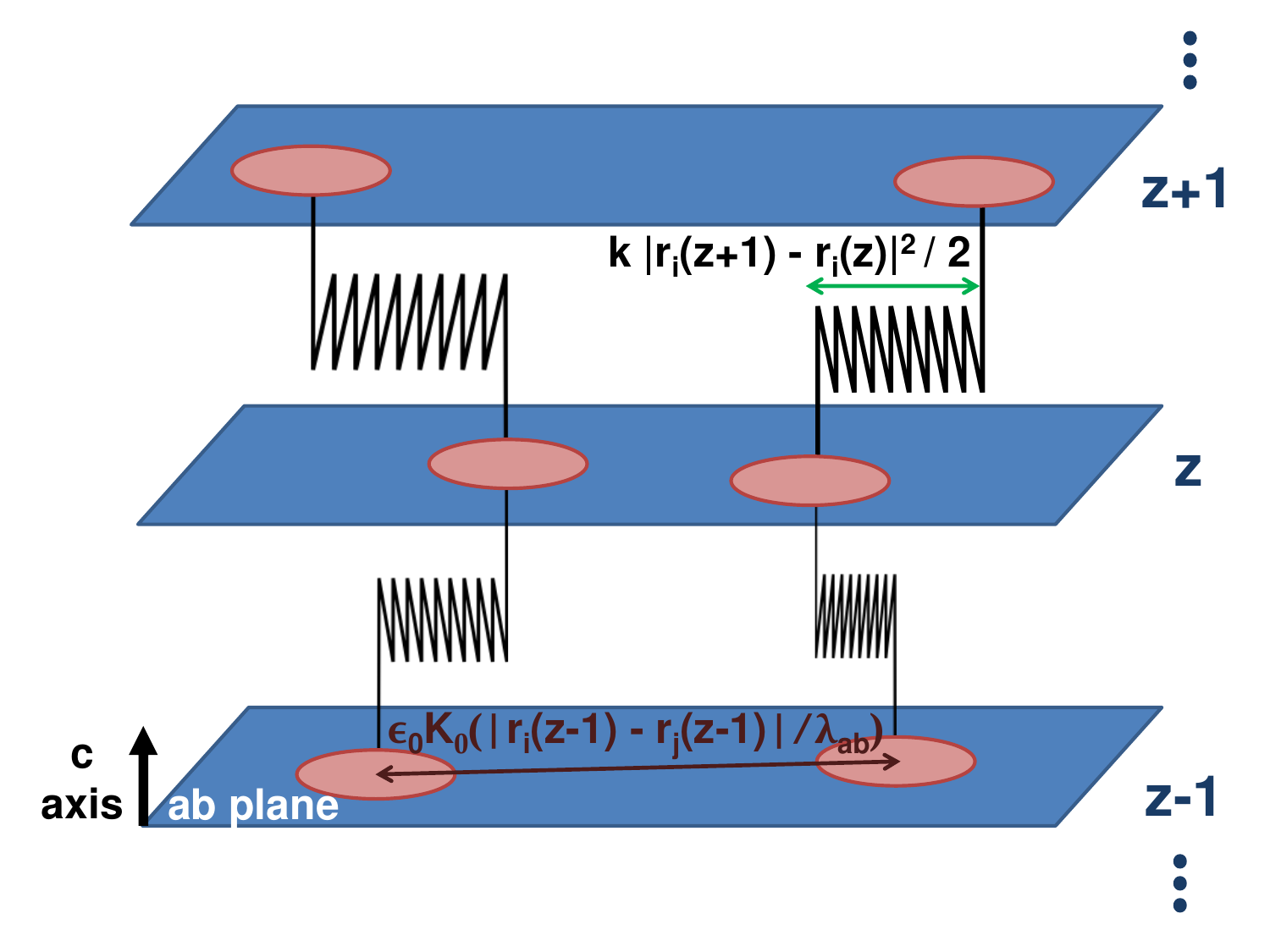}
\caption{Schematics of the model used in the Langevin dynamics simulations. Vortices are represented as stacks of particles (“pancakes”) coupled elastically along $z$ and interacting via London potentials within each layer.}
\label{Figure1}
\end{figure}


In order to address this question, here we perform numerical simulations of a model of a large number of interacting elastic lines in a three-dimensional medium with varying thicknesses.  This approach goes beyond the simple hydrodynamic theoretical analysis at thermal equilibrium and equilibrium simulations reported previously~\cite{Rumi2019}.  The simulation protocols followed here are  motivated by experiments revealing snapshots of structures that correspond to configurations frozen at a characteristic temperature where the dynamics are dramatically slowed down by the disorder of the host medium~\cite{CejasBolecek2016,Aragon2023,Puig2024}. This scenario is particularly relevant when cooling from a high-temperature liquid phase to a low-temperature solid phase. In the case of vortex matter, which serves as the experimental model system for our study, the system exhibits a liquid-to-solid first-order  transition upon cooling~\cite{Pastoriza1994,Zeldov1994,Dolz2014,Dolz2015}. We  reproduce this cooling process in our numerical simulations by slowly decreasing the temperature and passing through the melting temperature of the vortex system. This approach enables us to capture the effects of both, non-equilibrium and finite thickness, during the slow cooling of the system from a high-temperature liquid phase.

\section*{Methods}

Our simulations are motivated by experiments on highly layered vortex matter in Bi$_2$Sr$_2$CaCu$_2$O$_{8 + \delta}$ at low fields~\cite{Fasano1999,Fasano2003}. We model three-dimensional vortex lattices as collections of interacting elastic lines aligned with the  magnetic field applied along the \(c\)-axis. Each line consists of a stack of particles in consecutive layers, see schematics of Fig.\,\ref{Figure1}. Vortices are considered in the London limit, \(\lambda_{\rm ab}/\xi = \kappa \gg 1\), with \(\lambda_{\rm ab}\) the penetration depth and \(\xi\) the coherence length.

We perform Langevin dynamics simulations emulating the field-cooling process but neglecting pinning. During simulations the vortex structure is quenched from a high-temperature disordered liquid towards a low-temperature solid. To reproduce the available experimental data~\cite{Rumi2019,Besana2024}, we simulate 14,400 vortices with magnetic induction \(B \approx 15\)~G. The vortex density is controlled by the lattice spacing $a = 1.075\sqrt{\Phi_{0}/B} \sim 1.2$~$\mu$m with $\Phi_{0} = 2.07 \times 10^{-7}$~G\,cm$^2$ the flux quantum. In order to reproduce experimental data in Bi$_2$Sr$_2$CaCu$_2$O$_{8 + \delta}$, we take $\lambda_{\rm ab} \sim 0.5$~$\mu$m, the penetration depth at the temperature where pinning sets in, just below the vortex melting transition, for $B=15$\,G~\cite{CejasBolecek2016}.

For the numerical solution, we discretize space along the \(z\)-direction and represent each vortex line as a chain of particles. Within a vortex, a particle interacts elastically with its neighbors in adjacent layers, contributing
\begin{equation}
\frac{k}{2} \, |\mathbf{r}_i(z \pm 1) - \mathbf{r}_i(z)|^{2}   
\label{eq:resortes}
\end{equation}
to the total energy, where \(\mathbf{r}_i(z)\) is the in-plane position of particle \(i\) in layer \(z\), and \(k\) is an effective harmonic constant~\cite{Tauber1995,Assi2015}.  
Repulsive interactions between vortices in the same layer are described by the London interaction~\cite{Tinkham}:
\begin{equation}
V({\bf r}_i(z),{\bf r}_j(z)) = \epsilon_{0} K_{0} \left( \frac{|\mathbf{r}_i(z) - \mathbf{r}_j(z)|}{\lambda_{\rm ab}} \right),
\label{eq:londoninteraction}
\end{equation}
where \(K_0\) is the zeroth-order modified Bessel function of the second kind.

\begin{figure}[ttt]
\centering
\includegraphics[width=0.98\linewidth]{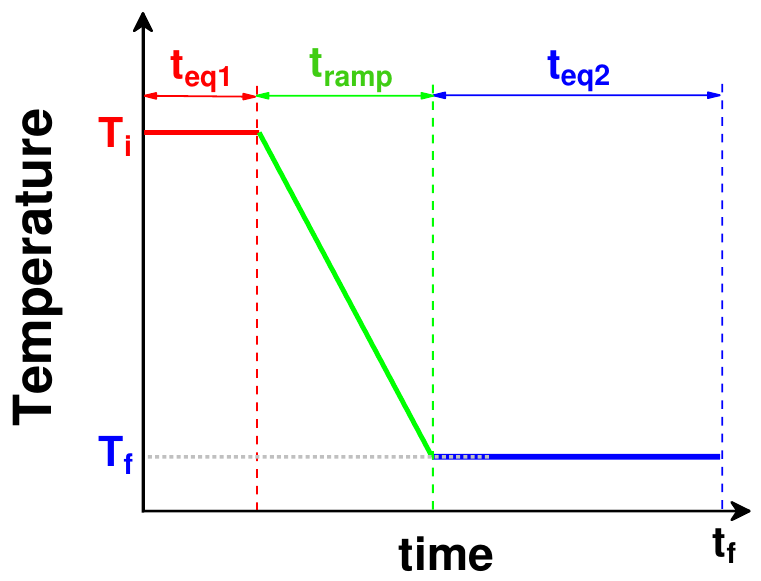}
\caption{Temperature-quench protocol used in simulations mimicking  field-cooling experimental conditions: Equilibration at $T_{\rm i}$, linear cooling during $t_{\rm ramp}$, and equilibration at $T_{\rm f}$.}
\label{Figure2}
\end{figure}

The overdamped Langevin dynamics of each particle is governed by
\begin{equation}
    \begin{aligned}
      \eta \frac{\partial \mathbf{r}_i(z)}{\partial t} &= 
      k \left[\mathbf{r}_{i}(z+\delta z)-2 \mathbf{r}_{i}(z) +\mathbf{r}_{i}(z-\delta z)\right]  \\
      &+ \sum_{j\neq i}   
      \frac{\epsilon_0}{\lambda_{\rm ab}}
      K_1\left(\frac{|{\bf r}_i-{\bf r}_j|}{\lambda_{\rm ab}}\right)
      \frac{{\bf r}_i-{\bf r}_j}{|{\bf r}_i-{\bf r}_j|}
     + \mathbf{f}_i(z).   
    \end{aligned}
    \label{eq:equationofmotion}
\end{equation}
where $\eta$ is an effective viscosity proportional to the Bardeen-Stephen viscosity, $\epsilon_0$ is the vortex line tension, and $K_1$ is the first-order modified Bessel function of the second kind. The stochastic force $\mathbf{f}_i$ models thermal fluctuations and satisfies 
$\langle \mathbf{f}_i(t) \rangle = 0$ and 
$\langle f^{\gamma}_i(t) f^{\gamma}_j(t') \rangle = 2 \eta (k_B T/\delta z) \delta_{ij} \delta_{\gamma,\gamma'} \delta(t-t')$, with $\gamma=x,y$  the cartesian components of the noise, and $\langle \dots \rangle$ denoting an ensemble average.  
Equation~\eqref{eq:equationofmotion} is integrated using a first-order Euler scheme with periodic boundary conditions in all three directions in a computational box of size $L_{x} \times L_{x}\times N_{z}$. This model captures the essential physics of vortex-vortex interactions and thermal effects, providing a reasonable approximation for the vortex structure in superconductors with $\kappa \gg 1$~\cite{Tauber1995}.

In the simulations of  Eq.\eqref{eq:equationofmotion} we consider all quantities dimensionless by measuring in-layer lengths in units of $\lambda_{\rm ab}$, lengths along the $c$-axis in units of $\Delta z$, time in units of $\eta \lambda^2_{\rm ab}/\epsilon_0$, energy in units of $\epsilon_0 \Delta z$, and temperature in units of $\epsilon_0 \Delta z / k_B$, with $k_B$ the Boltzmann constant. 
We treat $\Delta z$ as a tunable parameter in order to approximate the exact interactions between vortex segments of length $\Delta z$ using the London interaction.
We set the dimensionless spring constant $k\lambda_{\rm ab}^2/\epsilon_0=1$ such that in- and inter-layer interactions between particles are comparable at a distance $\lambda_{\rm ab}$ between particles. In what follows we refer exclusively to dimensionless variables.

The quenching protocol mimicking the experiments~\cite{Fasano1999,Fasano2003} starts from a vortex liquid of average separation $a$, equilibrated for $t_{\rm eq1}$ at $T_{\rm i} > T_{\rm m}$. The system is then cooled linearly over $t_{\rm ramp}$ to a final temperature $T_{\rm f} < T_{\rm m}$, followed by further equilibration during a time $t_{\rm eq2}$, see Fig.~\ref{Figure2}. 
For all simulations, we set $T_{\rm i} = 0.5$, $t_{\rm eq1} = 1.5 \times 10^4$, $t_{\rm eq2} = 5 \times 10^4$, and $t_{\rm ramp} = 2 \times 10^5$. To study the solid vortex phase after field-cooling, we choose $T_{\rm f} = 0.001$, sufficiently below $T_{\rm m}$. We have verified that the ramp is slow enough to ensure that structural properties, including the density of topological defects, are independent of the cooling rate.

With the aim of studying finite-thickness effects, we simulate vortex lattices with $N_z = 1$--35 layers, each containing 14,400 particles, averaging over four independent runs. The thickest system involves 504,000 particles, each interacting with $\sim 50$ in-plane neighbors via the London potential and with  two particles in adjacent layers via an elastic spring interaction, see Fig.\,\ref{Figure1}. Simulating slow temperature ramps requires $\sim 10^7$ evaluations of the London forces of Eq. \eqref{eq:equationofmotion} per time step, with typical ramps spanning over $2\times 10^7$ cooling steps, namely $\sim 10^{14}$ evaluations of pair interaction forces in a single simulation quench.

In order to accelerate the simulations, we developed a code that exploits GPGPU parallelism using a cell-list algorithm to efficiently compute the large number of London interactions~\cite{Anderson2008,Allen2017}. No pinning potential is included, which is justified because spatially uncorrelated disorder from point defects—as present in pristine Bi$_2$Sr$_2$CaCu$_2$O$_{8+\delta}$ samples—has mainly the effect of slowing down the out-of-equilibrium dynamics during the temperature quench~\cite{Aragon2023,Puig2024}. This effect can be interpreted as a temperature-dependent effective viscosity $\eta$ that rescales the timescale of the dynamics.

\section*{Results}

\begin{figure*}[ttt]
    \centering
    \includegraphics[width=0.82\linewidth]{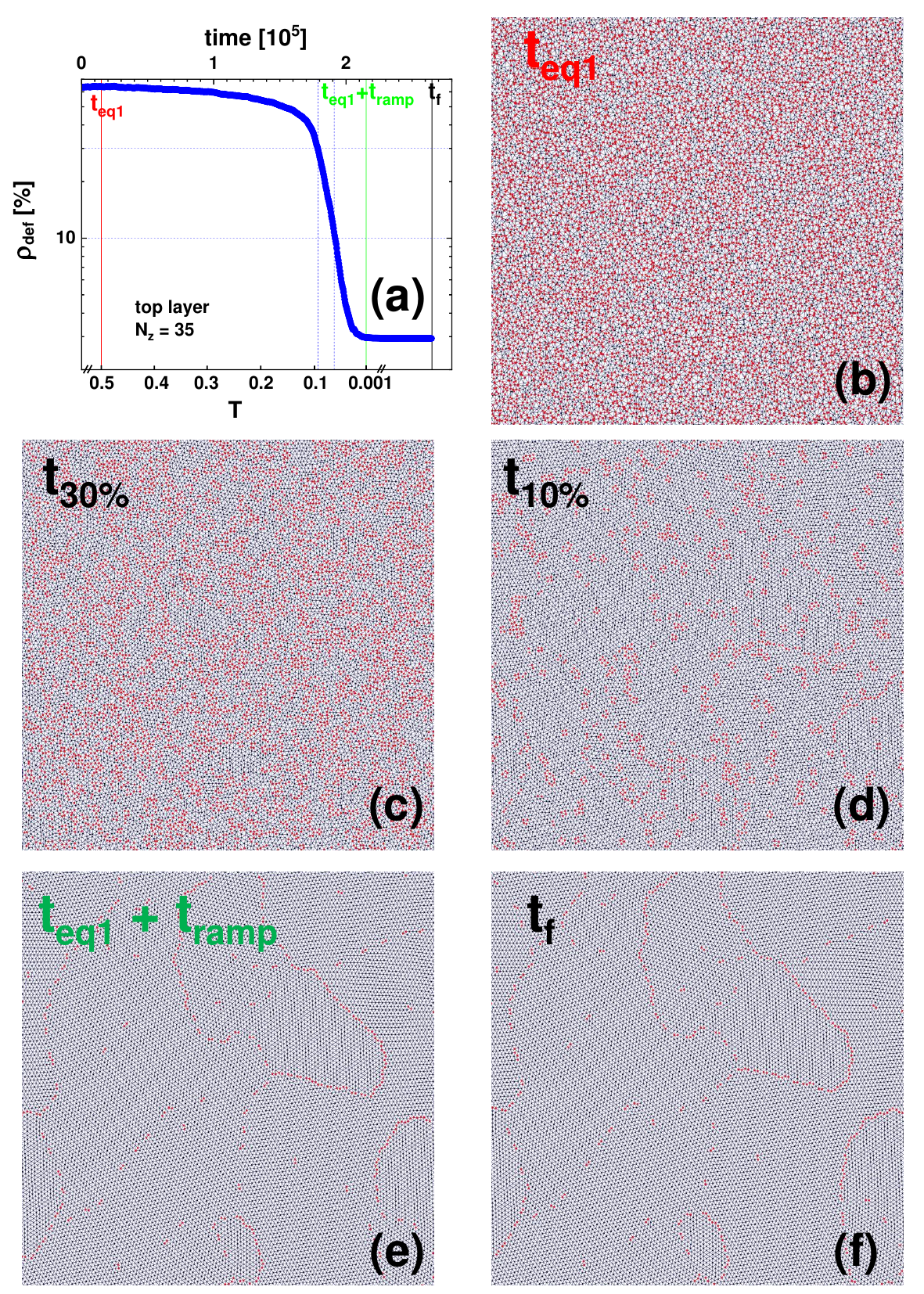}
    \caption{Vortex configurations and defect density during the quench for the thickest sample studied ($N_{z}=35$). 
    (a) Evolution of the defect density $\rho_{\rm def}$ as a function of time and temperature. 
    (b--f) Delaunay triangulations of the top layer at representative stages of the quench: 
    (b) equilibrated high-temperature liquid, 
    (c) at the temperature at which $\rho_{\rm def}\!\approx\!30\%$ and (d) $\!\approx\!10\%$, 
    (e) at the end of the ramp at $T_{\rm f}=0.001$, and 
    (f) final configuration after equilibration. 
    Non–sixfold vortices are marked in red.}
    \label{Figure3}
\end{figure*}

\begin{figure*}[ttt]
    \centering
    \includegraphics[width=0.9\linewidth]{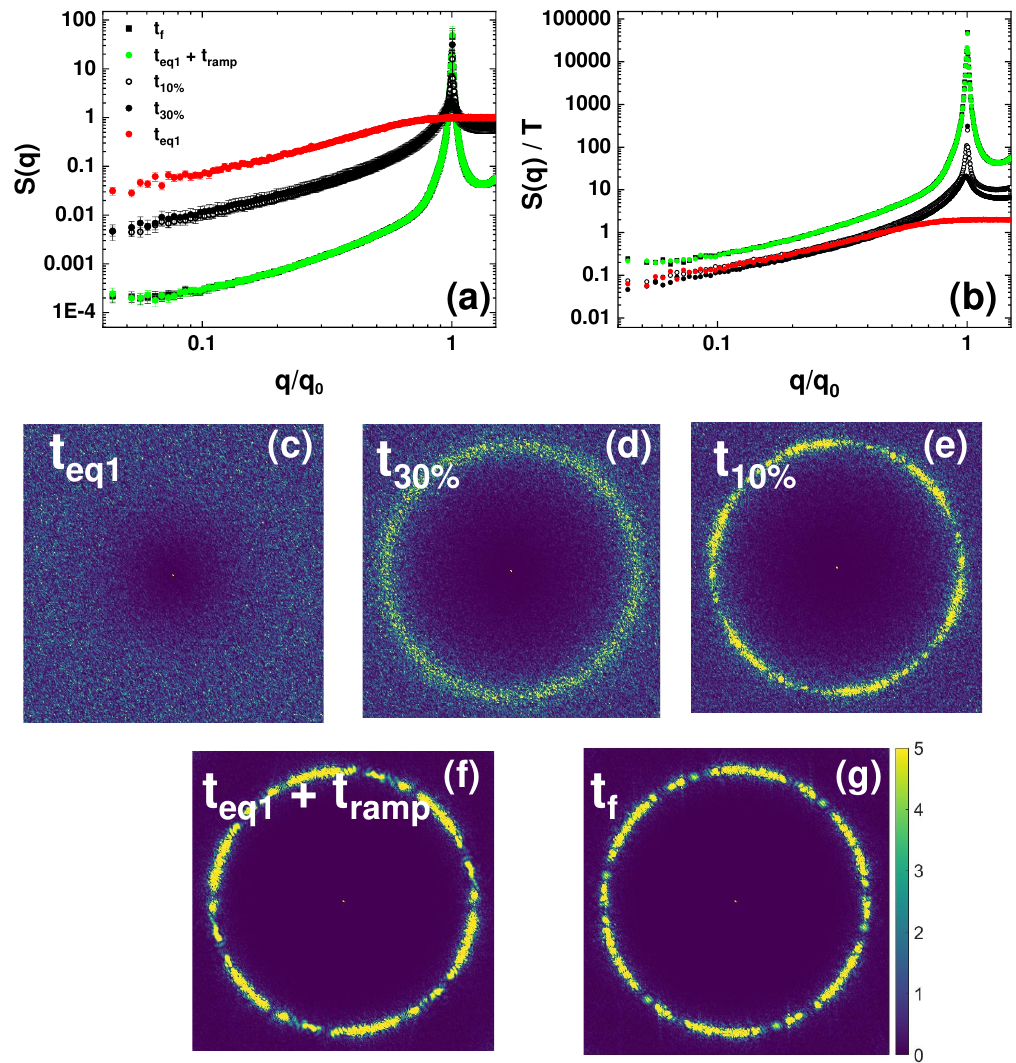}
    \caption{Structure factor of the vortex lattice for the thickest sample studied at different stages of the quenching protocol. 
    (a) Angularly averaged $S(q)$ at representative times/temperatures and (b) $S(q)/T$ for the same data. 
    (c--g) Two-dimensional structure factors corresponding to the indicated temperatures. 
    Results averaged over 35 layers and 4 independent simulations.}
    \label{Figure4}
\end{figure*}

Figure~\ref{Figure3} shows the vortex structure at different times during the quenching protocol for the thickest sample studied, $N_{z}=35$. Panel (a) displays the evolution of the defect density $\rho_{\rm def}$ as a function of time and temperature. Here, $\rho_{\rm def}$ is defined as the fraction of non-sixfold coordinated vortices in the top layer. The data correspond to the average over 4 independent simulations, with statistical fluctuations smaller than the symbol size. The non-sixfold coordinated vortices, forming topological defects typically associated with edge dislocations, are identified via Delaunay triangulation, as illustrated in panels (b)–(f). This triangulation identifies the first neighbors by finding the two closest vortices to each central vortex~\cite{Fasano2005}. In the figure, neighbor connections are shown with black lines, and non-sixfold coordinated vortices are highlighted in red.

Panel (b) of Fig.~\ref{Figure3} shows the Delaunay triangulation of the disordered vortex liquid at the initial temperature $T_{\rm i}=0.5$. At this temperature, the defect density is approximately 60\%. On cooling down to $T \sim 0.2$, $\rho_{\rm def}$ presents only a slight decrease. Upon further cooling, around $T \sim 0.1$ $\rho_{\rm def}$ exhibits a sudden drop, reaching roughly 3\% at $T = 0.001$, corresponding to a simulation time $t_{\rm ramp} + t_{\rm eq1} = 2.15 \times 10^{5}$. The system is then equilibrated for $t_{\rm eq2} = 0.5 \times 10^{5}$, during which $\rho_{\rm def}$ remains essentially constant. This final defect density is in quantitative agreement with experimental observations in Bi$_2$Sr$_2$CaCu$_2$O$_{8+\delta}$ vortex matter nucleated at $\sim 15$\,G and imaged at 4.2\,K after a field-cooling process from the liquid vortex phase~\cite{AragonSanchez2019}.

\begin{figure*}[ttt]
    \centering
    \includegraphics[width=0.95\linewidth]{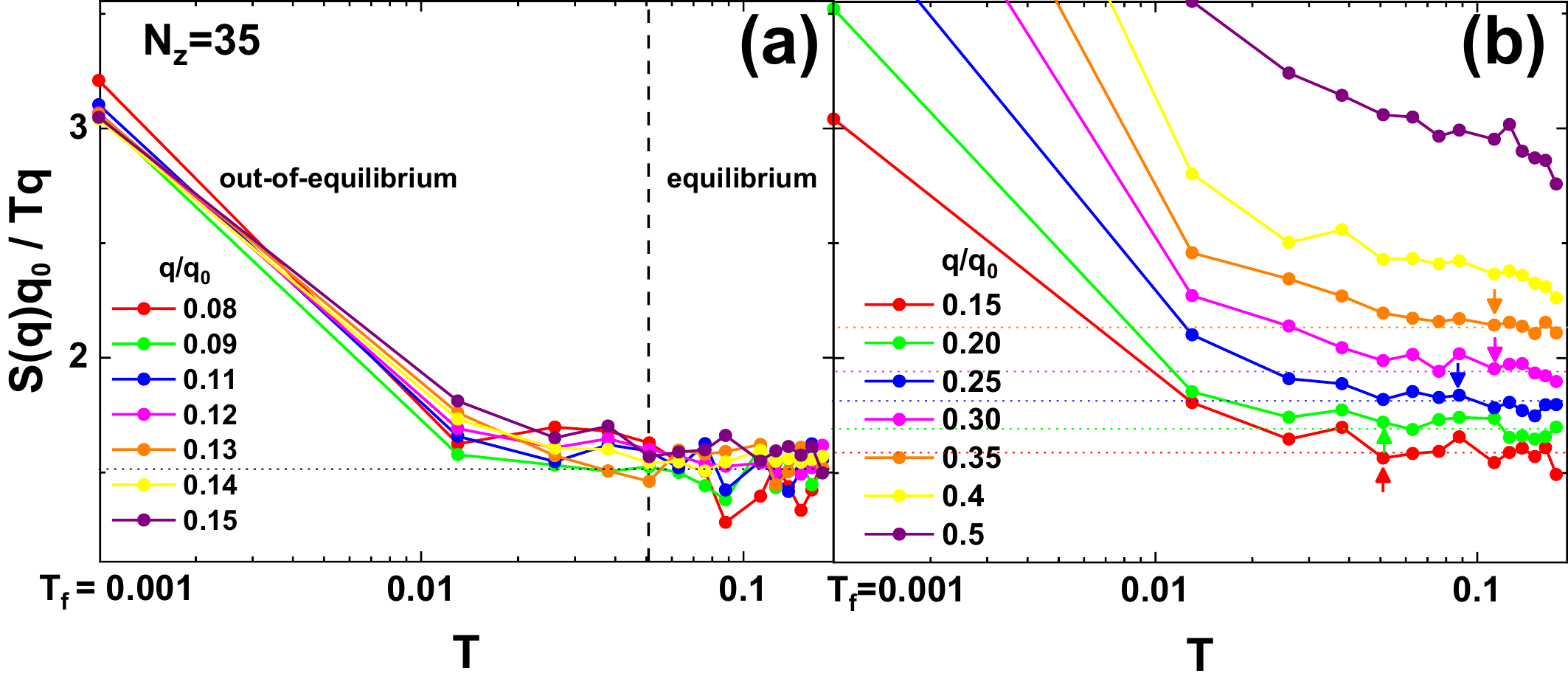}
    \caption{Angularly averaged structure factor normalized by $T$ and $q$ for the thickest sample studied with $N_{z}=35$, averaged over 4 simulation runs. (a) Low-$q/q_{0}$ data, where deviation from a constant signals loss of equilibrium upon cooling. 
    (b) Intermediate-$q/q_{0} \geq 0.15$ with arrows marking the temperature of departure from equilibrium for each wavevector.}
    \label{Figure5}
\end{figure*}

Panels (c) and (d) show the spatial distribution of defects corresponding to the sudden drop in $\rho_{\rm def}$ during cooling, at defect densities of 30\% and 10\%, respectively. Within this temperature range, the vortex structure begins to form crystallites that grow in size upon further cooling. The last two panels illustrate the structure at the final temperature $T_{\rm f} = 0.001$, first at the end of the temperature ramp ($t_{\rm eq1} + t_{\rm ramp}$, panel (e)) and then after an additional equilibration time $t_{\rm eq2}$ at this temperature (panel (f)). The final configuration at $t_{\rm f} = t_{\rm eq1} + t_{\rm ramp} + t_{\rm eq2}$ exhibits a polycrystalline structure with large grains, closely resembling the configuration observed at the end of the ramp.

Notably, the simulated defect density  at $T_{\rm f}$  and the simulated structures with grain boundaries surrounding crystallites of 300–400 vortices, agree quantitatively with experimental observations in Bi$_2$Sr$_2$CaCu$_2$O$_{8+\delta}$ vortex matter at 15\,G~\cite{AragonSanchez2019}. Since the simulations neglect disorder while real samples exhibit point pinning, this agreement further supports that including disorder primarily slows down the dynamics --and significantly increases computational time-- without altering the main equilibrium structural features~\cite{Aragon2023,Puig2024}.

The gradual ordering of the vortex structure upon cooling is also evident in the structure factor data shown in Fig.\,\ref{Figure4} for the thickest simulated sample with $N_{z} = 35$. Panels (a) and (b) display the angularly-averaged structure factor $S(q= \sqrt{q_x^2 + q_y^2})$ obtained by averaging the two-dimensional $S(q_x,q_y)$ data of panels (c)–(g) over the polar angle. Here, $S(q_x,q_y) = (1/N_z) \sum_{z=1}^{N_z} |\hat{\rho}(q_x,q_y,z)|^2$ is the squared modulus of the Fourier transform of the local density fluctuations of vortex positions, averaged over the $N_z$ layers. Data in panels (c)–(g) correspond to a single simulation realization, while the $S(q)$ data in panels (a) and (b) are averaged over four independent realizations, with the resulting dispersion in $S(q)$ being smaller than $10^{-4}$ even at low $q$ (see Appendix).

At the initial temperature, $S(q)$ is nearly structureless, but a Bragg peak emerges and grows in height upon cooling, as shown in Fig.\,\ref{Figure4}(a). The asymptotic value of $S(q)$ as $q \rightarrow 0$ becomes smaller on decreasing temperature, indicating that long-wavelength density fluctuations diminish on cooling. For the lowest simulated temperatures, when the defect density is small and nearly constant, the asymptotic $S(q)$ also stabilizes. The $S(q)$ curves corresponding to the end of the temperature ramp ($t_{\rm eq1} + t_{\rm ramp}$) and the final configuration after equilibration at $T_{\rm f}$ ($t_{\rm f}$) are nearly identical, confirming that the vortex configuration is effectively arrested on timescales of order $t_{\rm eq2}$.

Notably, for the thickest studied sample with $N_{z} = 35$, the final configuration shown in Fig.~\ref{Figure3} presents an $S(q)$ that decays algebraically as $q \rightarrow 0$, while its Bragg peak is very sharp. The algebraic decay indicates that long-wavelength vortex density fluctuations are strongly suppressed, whereas the sharp Bragg peak reflects the high crystalline order of the structure formed after the temperature ramp. Similar features are observed experimentally in the layered material Bi$_2$Sr$_2$CaCu$_2$O$_{8+\delta}$ at 15\,G~\cite{AragonSanchez2019,Besana2024}. The evolution of these features with sample thickness, or equivalently with the number of layers $N_{z}$, is analyzed in what follows to study finite-size effects.

\begin{figure*}[ttt]
    \centering
    \includegraphics[width=\linewidth]{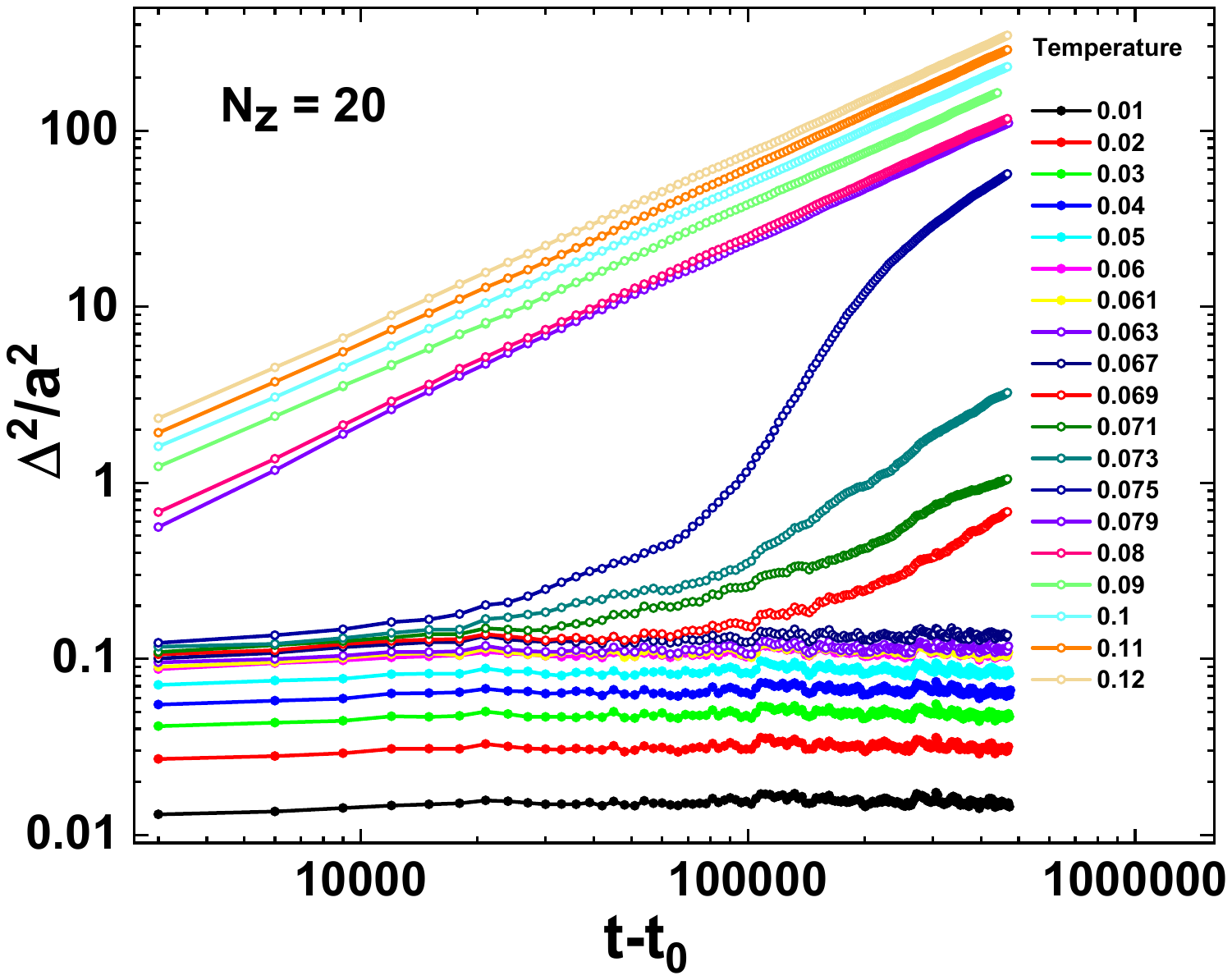}
    \caption{Diffusion curves $\Delta^{2}/a^{2}$ vs $(t-t_{0})$ for various cooling temperatures. 
    Results correspond to the $N_{z}=20$ sample, averaged over 4 simulations.}
    \label{Figure6}
\end{figure*}

Regarding the temperature dependence of the algebraic decay of $S(q)$ at small wave vectors, we recall that at thermal equilibrium $S(q) /T \propto  q$ since the elastic constants become non-dispersive in the $q\rightarrow 0$ limit~\cite{Rumi2019}. This follows from the equipartition theorem applied to elastic deformations of the vortex structure, which gives the longitudinal three-dimensional structure factor describing density fluctuations as
\begin{equation}
S(q_{x},q_{y},q_{z}) = \frac{n_{0} k_{\rm B} T q^{2}}{q^{2} c_{11}(q,q_z) + q_{z}^{2} c_{44}(q,q_z)},
\label{eq:sofq3d}
\end{equation}
where $n_{0}$ is the vortex density per unit area, $c_{11}$ the compression modulus, $c_{44}$ the tilt modulus, and $q_z$ the wavevector component perpendicular to the layers. This expression is valid for both, the liquid and  solid vortex phases~\cite{LeDoussal1990,marchetti1993,Rumi2019}. In the dilute vortex regime considered here, $c_{44}$ is approximately constant, while $c_{11}$ in the solid phase corresponds to the inverse of the isothermal compressibility.

Integrating Eq. \eqref{eq:sofq3d} over $q_z$, assuming  $c_{11}$ and $c_{44}$ are non-dispersive in $q_z$, and using periodic boundary conditions, yields the angularly averaged structure factor for a single layer~\cite{Blatter1994}:
\begin{equation}
S(q) = \tilde{A}(q) \, T q; \quad \text{with} \quad
\tilde{A}(q) = \frac{n_{0} k_{\rm B}}{\sqrt{c_{11}(q,0) c_{44}(q,0)}}.
\end{equation}
This result applies  at equilibrium to both,  the dilute  liquid and  solid vortex phases, considering the corresponding $c_{11}(q,0)$ for each phase. In the low-$q$ limit, where $c_{11}$ and $c_{44}$ are non-dispersive in the plane~\cite{Blatter1994}, we obtain $S(q) = \tilde{A}(0) \, T q$. Thus, rescaling the structure factor as $S(q)/T$ and observing a temperature-independent curve at low $q$ indicates thermal equilibrium. 
This criterion is expected to hold not only in the limit 
$q\to 0$, but more generally whenever 
$1/q < a_0$, i.e., in the regime where the elastic harmonic continuum description is valid. In the transition from low to intermediate $q$, the evolution of $S(q)/T$ is potentially $q$-dependent, $S(q)/T \propto {\tilde A}(q)q$,  due to the possible dispersion of the elastic constants $c_{11}(q)$ and $c_{44}(q)$ at wavevectors $q \cdot a_0 < 1$. In the limit  $q \to 0$ no dispersion of the elastic constants is expected and then $S(q)/T \propto q$.

Figure~\ref{Figure4}(b) illustrates this scaling at selected temperatures during the quenching protocol. Here, $S(q)$ is plotted as a function of $q/q_0$, with $q_0 \propto 1/a$ corresponding to the wavevector of the Bragg peak and inversely proportional to the average first-neighbor vortex spacing $a$. For temperatures between $T_i$ ($t = t_{\rm eq1}$) and $T \sim 0.07$, where $\rho_{\rm def} = 10$\%, the $S(q)/T$ curves collapse at small $q$, indicating that long-wavelength density fluctuations remain equilibrated during the temperature ramp. This collapse is exemplified by the three curves shown in Fig.~\ref{Figure4}(b) (open and filled black and red circles). On further cooling, the collapse no longer holds, and $S(q)/T$ increases abruptly, as seen for $T = T_{\rm f}$ at both $t_{\rm eq1} + t_{\rm ramp}$ and $t_{\rm f}$ (green circles and black squares). This behavior signals that the system globally falls out-of-equilibrium at a temperature between that of the sudden drop in the density of defects and the final simulation temperature.

\begin{figure}[ttt]
    \centering
    \includegraphics[width=1.05\linewidth]{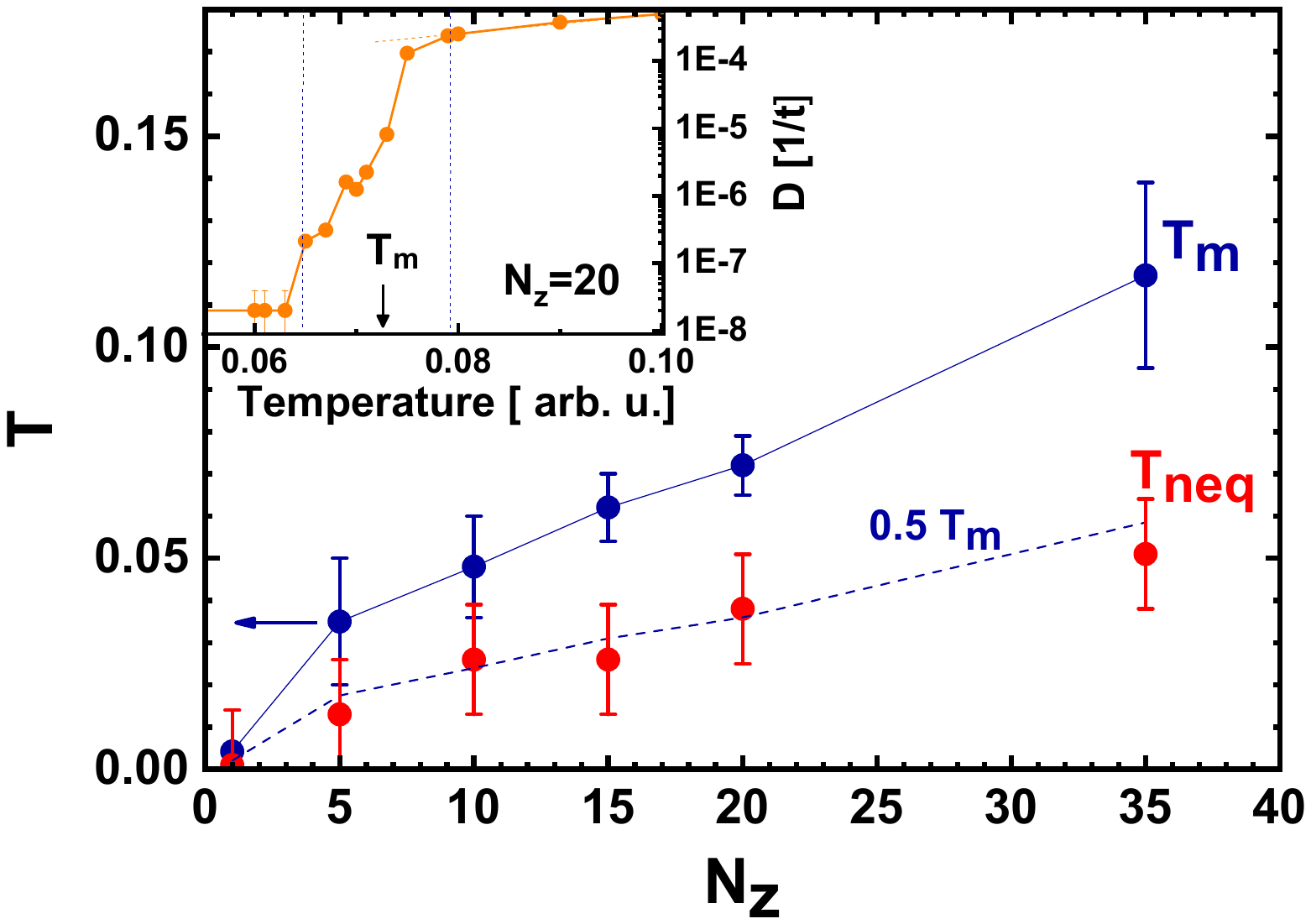}
    \caption{Melting temperature $T_{\rm m}$ and non-equilibrium temperature $T_{\rm neq}$ \textit{vs.} $N_{z}$. 
    The blue dotted line indicates $T_{\rm m}/2$, coinciding with $T_{\rm neq}$ within the error. 
    Insert: Diffusion coefficient $D$ as a function of temperature for $N_{z} = 20$. $T_{\rm m}$ is determined from the midpoint of the sharp drop; error bars reflect half the drop width.}
    \label{Figure7}
\end{figure}

\begin{figure*}[ttt]
\centering
\includegraphics[width=\linewidth]{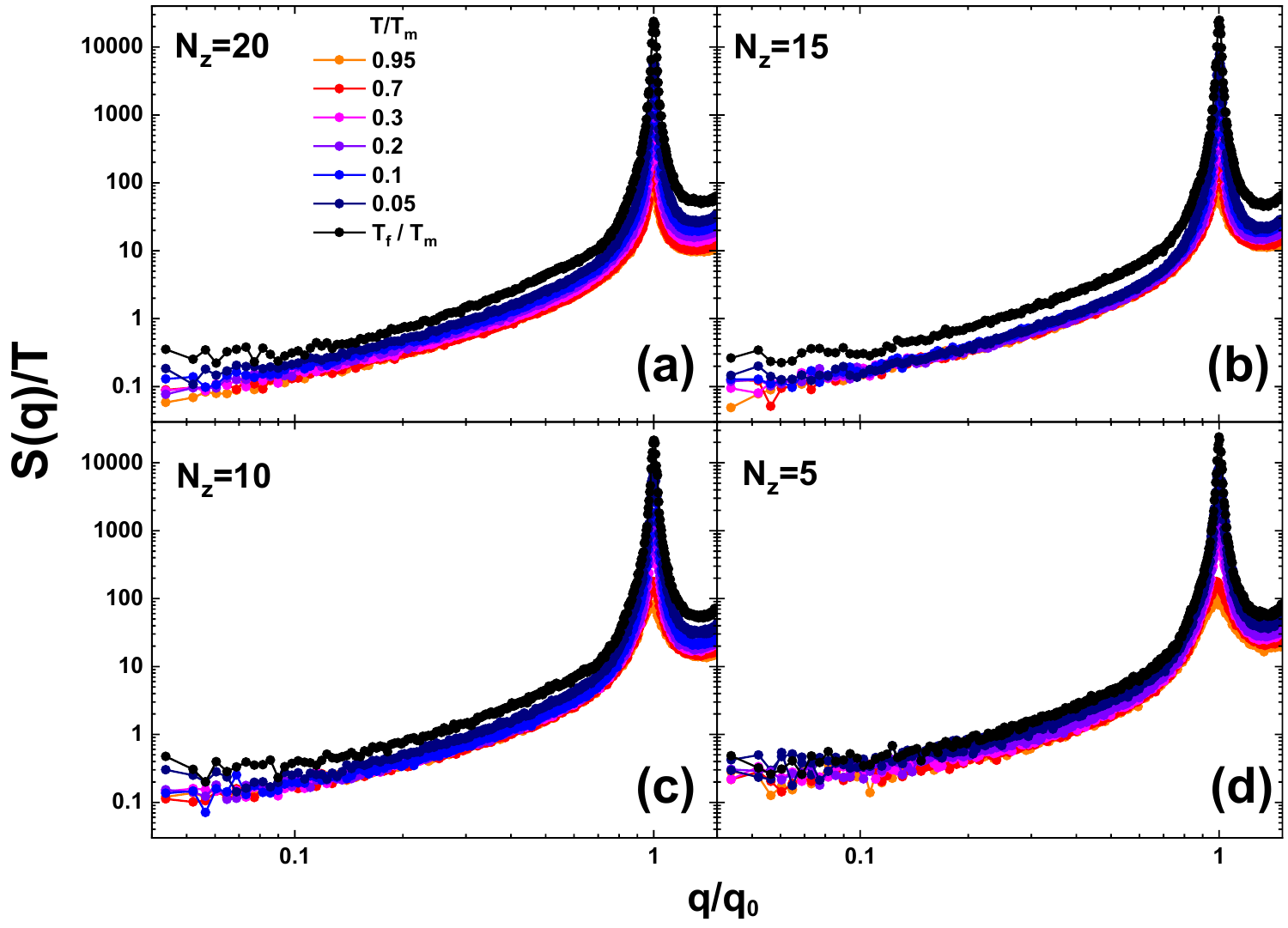}
\caption{Angularly averaged structure factor normalized by temperature, $S(q)/T$, at color-coded temperatures relative to $T_{\rm m}$. 
Data correspond to a single simulation for $N_{z}=$ (a) 20, (b) 15, (c) 10, and (d) 5.}
\label{Figure8}
\end{figure*}

\begin{figure*}[ttt]
\centering
\includegraphics[width=1.1\linewidth]{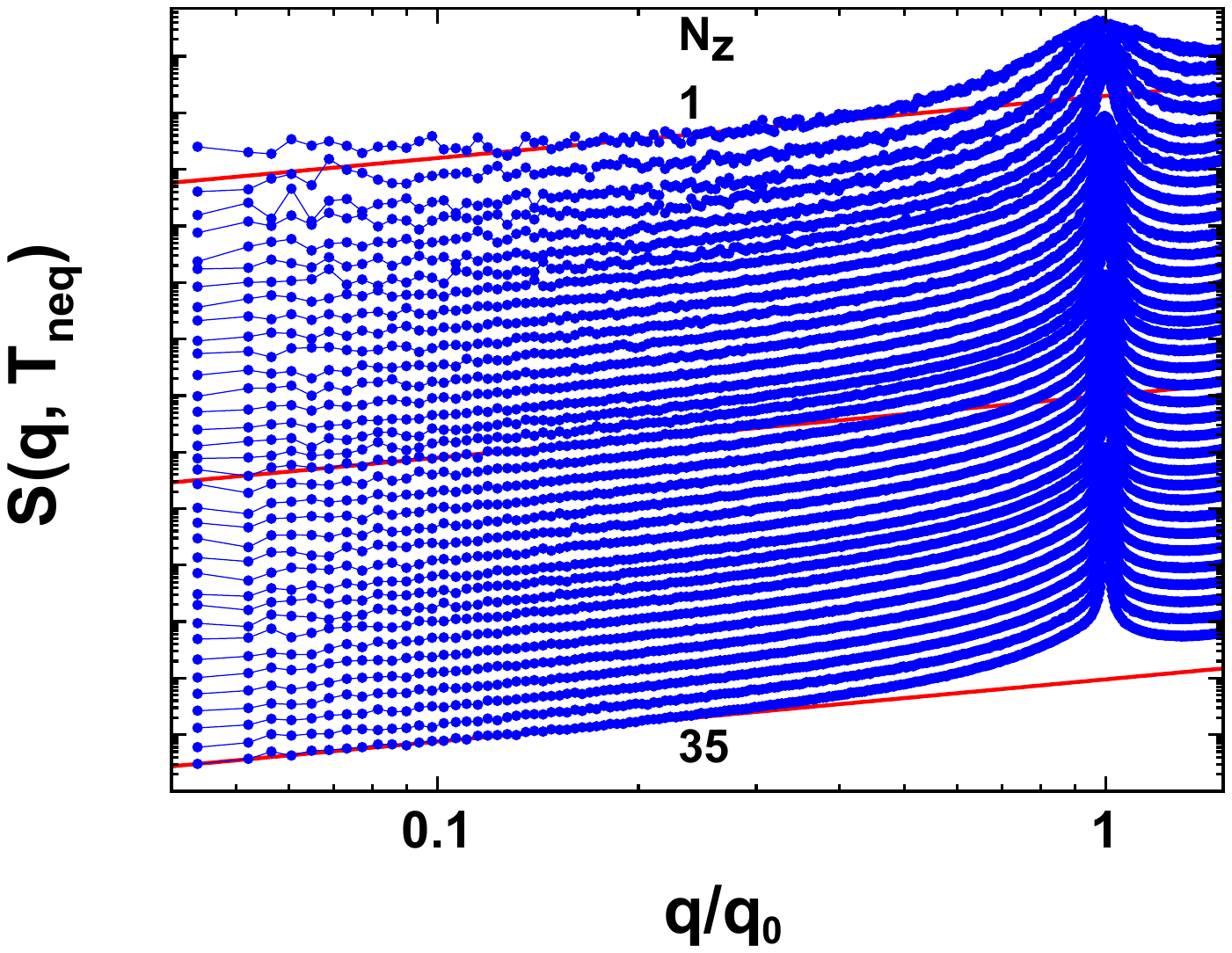}
\caption{Waterfall plot of the angularly averaged structure factor $S(q,T_{\rm neq})$ at the non-equilibrium temperature for all studied $N_z$ (1–35). Data averaged over 4 simulations. 
The bottom red line shows an algebraic fit of the $N_{z}=35$ data up to $q/q_{0}=0.22$ and its extrapolation, yielding $S(q) \propto (q/q_{0})^{1.1}$ at low $q$. 
The other two red lines correspond to the same curve vertically displaced for clarity.}
\label{Figure9}
\end{figure*}

Figure~\ref{Figure5} shows the temperature dependence of the normalized structure factor obtained in simulations, \(S(q)q_{0}/Tq = A(q)q_{0}\). The data correspond to the thickest sample studied, \(N_{z}=35\), averaged over four independent simulations, and are representative of the behavior observed for smaller \(N_{z}\).  For the following analysis, we compare the $A(q)$ data obtained in simulations with the $\tilde{A}(q)$ expected theoretically at equilibrium. For clarity, we separate the analysis into two regimes: (a) small $q$ where elastic constants are roughly non-dispersive; (b) larger $q$, where dispersion and eventually coupling between compression modes become relevant.  Panel (a) of Fig.~\ref{Figure5} shows that in the range \(0.08 \lesssim q/q_0 \lesssim 0.15\), \(A(q)\) remains nearly constant in temperature at high \(T\). Below  \(T \sim 0.05\), \(A(q)\) rises on decreasing temperature 
while retaining an approximate independence on \(q\). 
This phenomenology defines an empiric temperature $T_{\rm neq}$ where the systems falls out-of-equilibrium at large lengthscales greater than $\sim 10 a$ for the cooling rate adopted in our simulations. 
In contrast, panel (b) shows that for \(q/q_0 \gtrsim 0.2\), \(A(q)\) is temperature-independent only within a reduced high-temperature interval, which shrinks with increasing \(q\) (see arrows). In this temperature interval the system is thus at equilibrium. However, 
\(A(q)\) depends on $q$, indicating that dispersivity in the elastic constants becomes relevant. The temperature-dependent \(A(q)\) for high \(q\) mainly originates from coupling between compression modes inhibiting the application of the equipartition theorem. At the lowest \(q\) of panel (b) \(A(q)\) reveals the same phenomenology as in panel (a): the system falls out of equilibrium at a \(q\)-dependent characteristic temperature (see arrows).  This analysis provides quantitative evidence that the vortex configurations 
fall out-of-equilibrium at all lengthscales at an intermediate temperature $T \gtrsim T_{\rm neq}$ while performing the quenching protocol. This crossover,  between the high-temperature equilibrium structure and the low temperature out-of-equilibrium configuration, is relevant for comparison with experimental observations.

\begin{figure}[ttt]
\centering
\includegraphics[width=0.82\linewidth]{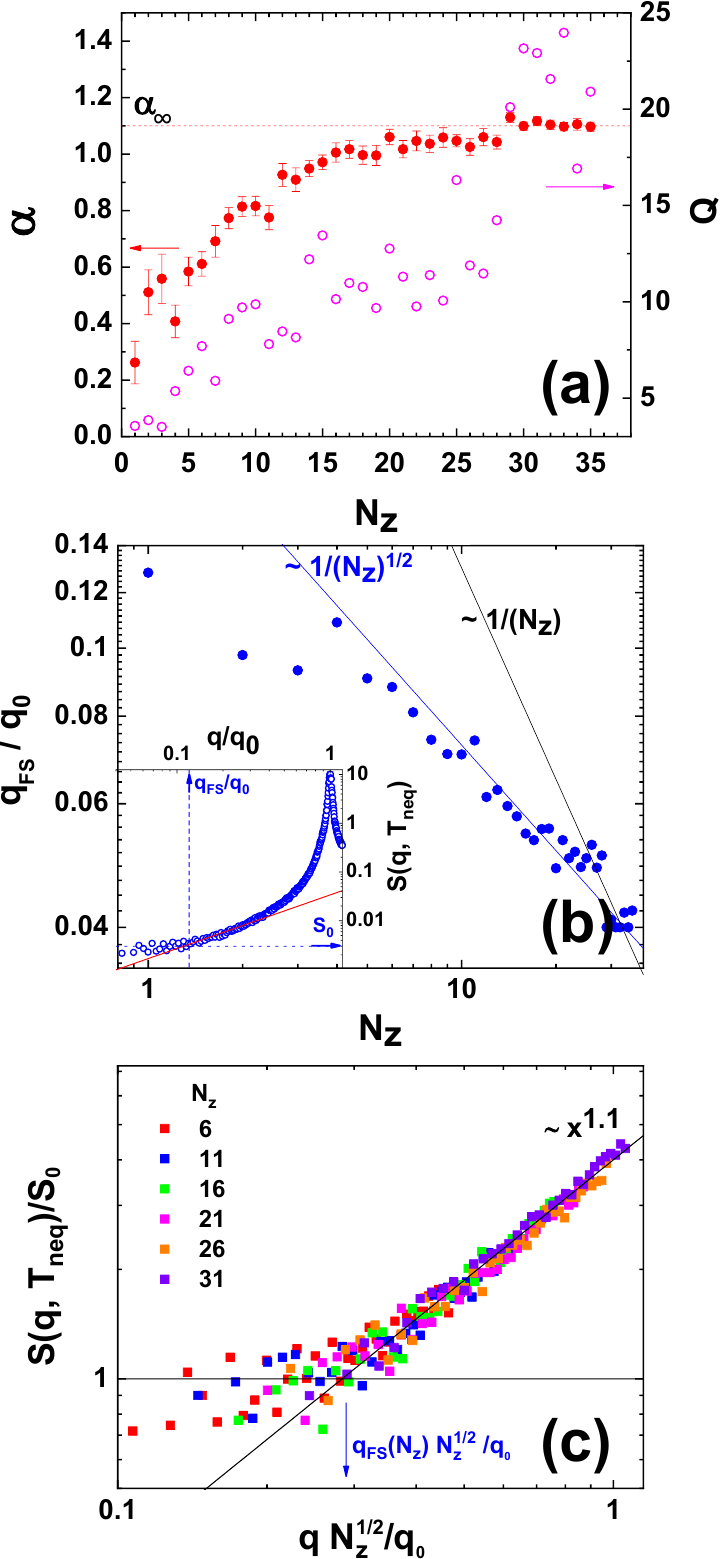}
\caption{(a) Effective exponent $\alpha$ vs $N_{z}$ obtained by fitting $S(q,T_{\rm neq})$ from the smallest $q$ up to $q_{\rm max}/q_{0}=0.22$ (red filled symbols), and the corresponding quality factor $Q$ (pink open symbols). 
(b) Crossover wavevector $q_{\rm FS}$ marking the onset of finite-size effects as a function of $N_z$. Insert: Determination of $q_{\rm FS}$ where $S(q)$ deviates from the algebraic behavior and approaches the stagnation value $S_0$, shown for $N_z=6$. 
(c) Structure factor normalized by $S_0$, $S(q,T_{\rm neq})/S_0$, plotted \textit{vs.} $q N_{z}^{1/2}$ for representative $N_z$. For $q N_{z}^{1/2} > q_{\rm FS} N_{z}^{1/2}$ the curves collapse onto an algebraic behavior with exponent 1.1; for smaller $q N_{z}^{1/2}$ the curves stagnate.}
\label{Figure10}
\end{figure}

The lowest $q/q_{0}$ data in Fig.~\ref{Figure5}(a) collapse onto a single curve at all temperatures, indicating that long-wavelength density fluctuations remain linear in $q$.
 This behavior is expected at equilibrium according to the low-$q$ theoretical expression for $A(q)$. Interestingly, for $q/q_{0} < 0.15$, the linear regime in $q$ persists even at low temperatures where the system is out of equilibrium. For  $q/q_{0} > 0.15$, the $\tilde{A}(q)q_0$ curves no longer collapse and instead shift to higher values. This shift reflects the $q$-dependence of $A(q)$ due to the dispersive nature of the elastic constants at larger $q$, signaling that density fluctuations enter a dispersive regime when $q/q_0 \gtrsim 0.15$.

The out-of-equilibrium crossover can also be characterized by the mean-squared displacement of vortex positions, $\Delta^{2}/a^{2}$, in simulations at constant temperature. To study this, we perform a cooling ramp with the same slope as before, but stop at the target temperature where the system evolves for a time $t_{0}$, after which we measure 
\begin{align}
\Delta^{2}(t,t_{0})/a^{2} &= (1/Na) \Sigma_{z=1}^{N_{z}}\Sigma_{i=1}^{N} [x_{i,z}(t) - x_{i,z}(t_{0})]^{2} \nonumber\\
&+ [y_{i,z}(t) - y_{i,z}(t_{0})]^{2}
\end{align}
where $N$ is the number of vortices per layer and $(x_{i,z},y_{i,z})$ denotes the position of vortex $i$ in layer $z$. The equilibration time $t_{0}$ is chosen such that $\Delta^{2}(t,t_{0})/a^{2}$ depends only on the difference $t-t_{0}$ in the temperature range studied.  
Figure~\ref{Figure6} shows $\Delta^{2}/a^{2}$ as a function of simulation time for $N_{z}=20$, a representative case for all the $N_{z}$ studied. At high temperatures, $\Delta^{2}/a^{2}$ increases algebraically with time. At intermediate temperatures, a dynamic change occurs where fluctuations drop sharply and exhibit nonlinear behavior. 
At low temperatures, $\Delta^{2}/a^{2}$ is strongly suppressed and remains nearly constant. For $N_{z}=20$, this crossover occurs between $T \approx 0.08$ and $0.07$, with a slight dependence on $N_{z}$, as discussed later.

We recall that in the liquid phase, a normal diffusion regime with $\Delta^{2}(t,t_{0})/a^{2} \sim (t-t_{0})$ is expected, as observed in Fig.\,\ref{Figure6} for $T > 0.079$. The observations for $T < 0.069$ are consistent with expectations for a solid phase with $\Delta^{2}(t,t_{0})/a^{2}$ small and nearly constant over time. In the intermediate temperature range, $0.069 < T < 0.079$, the system exhibits a sudden change in dynamic behavior, with fluctuations transitioning from liquid-like at high temperatures to solid-like at low temperatures. Within this range, our quenching protocol effectively captures the melting process, modeling the first-order melting transition observed in layered vortex matter at high temperatures~\cite{Pastoriza1994,Zeldov1994}. Hence, our simulation protocol reasonably mimics the experimental field-cooling procedure used to image vortex structures at low fields~\cite{Fasano1999}.

The melting transition can be identified by computing the diffusion coefficient $D$ of the vortex structure as a function of temperature. This coefficient is obtained by fitting the asymptotic linear dependence of the mean-square displacement, $\Delta^{2}/a^{2} = D \cdot (t-t_{0})$ for $t-t_{0} \rightarrow \infty$. Applying this procedure to the data of Fig.\,\ref{Figure6} for $N_{z}=20$, we obtain the temperature dependence of $D$ shown in the insert to Fig.\,\ref{Figure7}. During cooling, $D$ exhibits a sudden drop from a finite high-temperature value to a value below the simulation resolution ($\sim 5 \cdot 10^{-8}$) for $T < 0.064$, signaling the melting of the vortex structure. We define the melting temperature $T_{\rm m}$ as the midpoint of the temperature range over which $D$ decreases from 90\% to 10\% of its high-temperature value (dotted blue lines and black arrow in the inset). The position of this jump, and thus $T_{\rm m}$, depends strongly on the sample thickness. Repeating this procedure for various $N_{z}$ yields the results shown in the main panel of Fig.\,\ref{Figure7}. The melting temperature determined in this way is in agreement with the Hansen-Verlet criterion for vortex melting~\cite{Cornaglia1999}, see Appendix.

For $N_{z} \geq 5$, $T_{\rm m}$ increases roughly linearly with thickness.  In order to understand this thickness-dependent melting temperature, we recall that melting is controlled by shear modes that 
start in the solid phase where the three-dimensional shear modulus $c_{66}$ is finite and that these modes do not change the local density.  In the solid phase, when the shear elastic energy equals the tilt energy, the system undergoes a dimensional crossover for shear deformations between a three to a two-dimensional behavior.
Equating the shear elastic energy  $E_{shear}\sim c_{66} q^2$ with the tilt  energy  $E_{tilt} \sim c_{44} q_z^2$ for a  harmonic elastic deformation with wave vectors  $q$ for shear and $q_z$ for tilting, allows to obtain the characteristic length in the $z$-direction 
\begin{equation}
\xi_{sh}(q) \approx q^{-1}\sqrt{\frac{c_{44}}{c_{66}}} \propto q^{-1}\sqrt{\frac{k}{c_{66}}}.
\label{eq:xish}
\end{equation} 
When this length becomes larger than the thickness of the system (proportional to $N_{z}$), there is a dimensional crossover for shear deformations between a three to a two-dimensional behavior of shear fluctuations. For fixed $c_{66}$ and $k$ values, short wavelength (large $q$) shear deformations will be in a three-dimensional regime whereas large wavelength (short $q$) shear deformations will be in a two-dimensional regime. Since melting is controlled by the large wavelength shear modes, we apply the two-dimensional  
Lindemann melting criterion with an effective two-dimensional shear modulus $N_z c_{66}$. We thus obtain the estimate for the melting temperature  
\begin{equation}
c_{\rm Li}^2 a^2 \sim \frac{T_m}{c_{66} N_z},    
\end{equation}  
with $c_{\rm Li} \approx 0.15$~\cite{Nattermann2000}. This implies $T_m \propto N_z$ as observed in the data of Fig.~\ref{Figure7}. Nevertheless, for very thick samples ($N_z$ large) a melting temperature independent of thickness is expected~\cite{Nattermann2000}.

Regarding hyperuniformity, this property is governed by longitudinal fluctuations controlled by compression deformations of the vortex structure. These deformations are 
 characterized by a different length scale in the $z$-direction. Analogously to the previous analysis, we can thus obtain the characteristic length in the z-direction for compression deformations
\begin{equation}
\xi(q) \sim q^{-1} \sqrt{\frac{c_{44}}{c_{11}}}.
\label{eq:xi}
\end{equation}
As will be discussed later, $\xi < \xi_{sh}$, indicating the three- to two-dimensional crossover of compression deformations will occur at smaller $q$ compared to the crossover in shear deformations.

Irrespective of these crossovers, for all the studied thicknesses   the vortex structure undergoes a relatively sharp liquid-to-solid transition at $T_{\rm i} < T_{\rm m} < T_{\rm f}$. The final configuration at $T_{\rm f}$ is out-of-equilibrium since, as discussed previously, the system falls out of equilibrium at an intermediate temperature $T_{\rm neq} < T_{\rm m}$. Using the same analysis as in Fig.~\ref{Figure5}(a), we extract $T_{\rm neq}$ for various values of $N_{\rm z}$ (red points in Fig.~\ref{Figure7}).  
Equilibration time increases with decreasing wavevector $q$. Thus, we empirically define $T_{\rm neq}$ as the temperature at which local equilibration is achieved up to length scales $\sim 10a$, i.e., for modes with $(q/q_0) < 0.1$.
Remarkably, we find $T_{\rm neq} \sim 0.5 T_{\rm m}$, suggesting that both characteristic temperatures are governed by the same $N_z$-dependent energy scale. Since our simulations neglect the effect of disorder, the estimated $T_{\rm neq}$ should be regarded as a lower bound for more realistic models including pinning.

In experimental cooling protocols for vortex imaging,~\cite{Besana2024} the solid vortex structure becomes frozen at length scales of the order of $a$ near the irreversibility temperature, when pinning effects start to dominate~\cite{CejasBolecek2016}. Upon further cooling, the frozen vortex structure is out-of-equilibrium. Therefore, in order to compare our disorder-free simulations with experiments in real samples with weak pinning, it is natural to consider simulated configurations at temperatures close to $T_{\rm neq}$ where the system effectively falls out of equilibrium.

Figure\,\ref{Figure9} shows the angularly averaged structure factor at the corresponding $T_{\rm neq}$ for each studied $N_{z}$ between 1 and 35. The data of $S(q,T_{\rm neq})$ were obtained by averaging over 4 independent simulations for each $N_{z}$ value. For the thickest samples with $N_{z}>30$, the low-$q/q_{0}$ behavior is well described by $S(q,T_{\rm neq}) \propto (q/q_{0})^{\alpha}$ with $\alpha = (1.1 \pm 0.02) = \alpha_{\infty}$, see Fig.\,\ref{Figure10}(a). This algebraic scaling indicates that the simulated vortex system is hyperuniform, in agreement with theoretical predictions for sufficiently thick samples~\cite{Rumi2019} and with experimental observations in Bi$_2$Sr$_2$CaCu$_2$O$_{8+\delta}$.~\cite{Rumi2019,Puig2022,Besana2024} Although hyperuniformity is an asymptotic  property in the $q \rightarrow 0$ limit,  the $S(q,T_{\rm neq})$ curves display an algebraic scaling up to $q_{\rm max}/q_{0} \sim 0.22$, independent of $N_{z}$ for the range studied. For larger $q/q_{0}$ the structure factor is dominated by the dispersive character of the elastic constants.

For $N_{z}<30$, the algebraic behavior in 
Fig.\,\ref{Figure9} is observed only within a restricted $q/q_{0}$ window, bounded below by an $N_{z}$-dependent cutoff and above by $q_{\rm max}/q_{0}=0.22$ for all thicknesses. Nevertheless, if a power-law fit of $S(q,T_{\rm neq})$ is forced between the smallest simulated $q$ value and $q_{\rm max}$, the resulting effective exponent $\alpha$ decreases systematically with decreasing $N_{z}$, see Fig.\,\ref{Figure10}(a). The quality factor of the fits, 
that we heuristically define as $Q=\exp(-\Delta \alpha / \alpha)/\chi^{2}$, 
increases with sample thickness. Here $\Delta \alpha$ is the fitting error on $\alpha$ and $\chi^{2}$ the goodness-of-fit parameter. Notably, $Q$ decreases significantly for small $N_{z}$. A closer inspection of the data in Fig.\,\ref{Figure9} shows that this reduction in $Q$ originates from the tendency of $S(q,T_{\rm neq})$ to saturate at small $q/q_{0}$ when $N_{z}<30$.

The saturation highlighted in the insert of Fig.\, \ref{Figure10}(b) can be understood as a finite-thickness crossover effect, governed by a $N_{z}$-dependent wavevector $q_{FS}$ that marks the onset of saturation as $q$ decreases at fixed $N_{z}$. A simultaneous fit of the growing exponent of $S(q)$ and $q_{FS}$ from the data is not feasible, particularly when $q_{FS} \ll q_0 $, since $S(q,T_{\rm neq})$ becomes increasingly noisy in the $q/q_{0} \to 0$ limit. Alternatively, in order to estimate $q_{FS}$ we assume that the $S(q)$ data are well described by

\begin{equation}
S(q) \sim F(q,q_{*}) = q^{\alpha_{\infty}} G(q/q_{*}),
\end{equation}
with $G(x) = G_{1}x^{-\alpha_{\infty}}$ for $x<1$ and $G(x) = G_{1}$ for $x>1$, where $G_{1}$ is a constant. We then compute the mean least-squares residual
\begin{equation}
R(q_{*},\gamma) = \sum_{q=q_{\rm min}}^{q_{\rm max}} \left[ F(q,q_{*}) - q^{\gamma} \right]^{2},
\end{equation}
and minimize it with respect to $q_{*}$ and $\gamma$ in the range $q_{\rm min}=2\pi/L_{x}$, the smallest accessible wave vector, and $q_{\rm max}/q_{0}=0.22$. 
This procedure yields a set of pairs of $\gamma$ and $q_{*}$. We infer $q_{FS}$ as the particular value of $q_{*}$ when $\gamma$  equals
$\alpha$ of Fig.\,\ref{Figure10}(a). Following this procedure for all the studied $N_{z}$, we obtain the thickness-dependent $q_{FS}$ shown in Fig.\,\ref{Figure10}(b). This magnitude follows the scaling law $q_{FS} \sim N_{z}^{-1/2}$ for sufficiently large $N_{z}$.

We can further test the previous heuristic procedure for estimating $q_{FS}$ by rescaling the structure factor data for different $N_{z}$ onto a single master curve. Figure\,\ref{Figure10}(c) shows $S(q,T_{\rm neq})/S_{0}$ plotted as a function of $q/N_{z}^{1/2}$, where $S_{0}$ is the low-$q$ saturation value (see insert to Fig.\,\ref{Figure10}(b)). The successful collapse of the curves confirms the scaling form 
\begin{equation}
S(q,T_{\rm neq}) \approx q^{\alpha_\infty} \, G(q N_{z}^{1/2}),
\end{equation}
as anticipated.

\section*{Discussion}

It is interesting to note that our system can exhibit topological defects and hyperuniformity simultaneously, whereas in some systems topological defects or imperfections are responsible for small-$q$
deviations from hyperuniformity~\cite{KimTorquato2018PRB97,ChenZhengJiao2021PRB104}.
In the vortex system, the relationship between defects and small-$q$ deviations is more subtle: At equilibrium, hyperuniformity is predicted even in the vortex liquid state with zero shear modulus, despite having a high density of dislocations. This is the result of hyperuniformity in superconducting vortex systems not being controlled by variations in shear modes but primarily by tilt and compression modes.~\cite{Rumi2019}
This observation suggests that the mere presence of defects does not necessarily destroy in-plane hyperuniformity in interacting three-dimensional elastic-line structures. However, their role in out-of-equilibrium hyperuniformity, and in particular their possible impact on the finite-thickness crossover, remains an open and intriguing question.

The observed power-law decay of $q_{FS}$ with increasing $N_z$ can be qualitatively understood as follows. At sufficiently small in-plane wave vectors, the density fluctuations of both, the liquid and solid vortex phases, follow the three-dimensional equilibrium structure given by Eq.~\eqref{eq:sofq3d}. Fourier transforming $S(q_x,q_y,q_z)$ only in $q_z$, and imposing periodic boundary conditions along the $z$ direction, yields the spatial correlation function~\cite{marchetti1993}  
\begin{equation}
S({\bf q},z_1-z_2) = \frac{n_0 k_B T}{c_{11} \,\xi(q)} \,
e^{-|z_1-z_2|/\xi(q)},
\end{equation}  
where the $q$-dependent compression correlation length along $z$ is given by Eq.~\eqref{eq:xi}. This expression shows that modes with smaller in-plane wave vector $q$ develop longer correlation lengths along $z$. Consequently, for sufficiently small $q$, finite-thickness effects are theoretically expected when $q \approx \tilde{q}_{FS}$, such that $\xi(\tilde{q}_{FS}) \sim L$, with $L \propto N_z$ the system thickness. The corresponding crossover wave vector is then  
\begin{equation}
\tilde{q}_{FS} \propto \frac{1}{N_z}\sqrt{\frac{c_{44}}{c_{11}}},
\label{eq:qfs}
\end{equation}  
where both $c_{44}$ and $c_{11}$ are non-dispersive at small $q$ and $q_z$.  It is worth noting that this finite-thickness crossover wave vector for density fluctuations  is analogous to the crossover $q_{\rm shear} \propto N_z^{-1} \sqrt{c_{44}/c_{66}}$ for transverse constant-density fluctuations, obtained by setting $\xi_{\rm sh} \sim L$ in Eq.~\eqref{eq:xish}, which controls the effective dimensionality of the melting transition. 
Interestingly, since $c_{11}\gg c_{66}$, one finds $q_{\rm shear}>{\tilde q}_{FS}$, indicating that shear modes cross over from three- to two-dimensional fluctuations at shorter length scales than compression modes. As a result, hyperuniformity—which cannot occur at equilibrium in two-dimensional systems with short-range interactions—can coexist with two-dimensional long-wavelength shear fluctuations in the vortex solid, within the regime ${\tilde q}_{FS}<q<q_{\rm shear}$. This scenario is consistent with the thickness-dependent melting temperature shown in Fig.~\ref{Figure7}.

This prediction of $\tilde{q}_{FS} \sim N_z^{-1}$, although capturing a power-law crossover, decays faster than the behavior observed in simulations $q_{FS} \sim N_z^{-1/2}$, see Fig.~\ref{Figure10}. The origin of this discrepancy lies in the fact that Eq.~\eqref{eq:qfs} was derived under the assumption of global thermal equilibrium, whereas the results in Fig.~\ref{Figure10} correspond to $T_{\rm neq}$, where only modes with wave vectors larger than $(q/q_0)\sim 0.1$ achieve local equilibration. Indeed, Fig.~\ref{Figure10}(b) shows that the scaling $q_{FS} \sim N_z^{-1/2}$ holds for $q_{FS}/q_0 < 0.1$, indicating that the observed $q_{FS}$ is a non-equilibrium crossover. The apparent saturation of $S(q)$ for $q < q_{FS}$ thus reflects the slow dynamics of long-wavelength modes that can not equilibrate for the simulation cooling rate. This also explains why the predicted $\tilde{q}_{FS}$ is smaller than the empirical values extracted from simulations $q_{FS}$. These results highlight that finite-thickness effects also manifest in out-of-equilibrium conditions, thereby destroying hyperuniform patterns upon cooling.

While we do not yet have a fully quantitative derivation for the $N_z^{-1/2}$ scaling on $q_{FS}$, we can offer some physical intuition on why it deviates from the equilibrium $N_z^{-1}$ scaling.
At equilibrium, the scaling $q_{FS} \sim N_z^{-1}$ follows from the three-dimensional structure factor derived from the equipartition theorem, see
Eq. \eqref{eq:sofq3d},
when balancing the terms in the denominator $c_{11} q^2 \sim c_{44} q_z^2$ and considering the constraint $q_z \sim 1/N_z$. These considerations yields the scaling
$q_{FS} = \sqrt{c_{44}/c_{11}}/N_z \sim 1/N_z$.
Out of equilibrium, the equipartition theorem no longer holds. On decreasing $q$  the density modes decay more slowly and retain a stronger memory of the high-temperature initial conditions. This introduces a $q$-dependent distortion in $S(q, t)$. Assuming that tilt modes equilibrate faster than compression modes, 
this memory effect can be described by an effective temperature $T_{\text{eff}}(q, t) > T(t)$ affecting primarily compression modes, where $T(t)$ describes the evolution of temperature during the cooling process. This effectively modifies the compression moduli such that $c_{11} \to c'_{11}(q, t)$ while $c_{44}$ remains temperature and time independent as in equilibrium. The new crossover wavevector would then satisfy
$$q'_{FS} \approx \frac{\sqrt{c_{44}/c'_{11}(q'_{FS}, t)}}{N_z}$$
If the effective elastic constant scales as $c'_{11} \sim q^2$ during this relaxation process, we naturally arrive at the observed $q'_{FS} \sim N_z^{-1/2}$ scaling. While this description is a preliminary heuristic argument, it is physically consistent with the observed distortions in $S(q,t,T(t))$ relative to its equilibrium value at each temperature $T(t)$  during the cooling process, $S_{\rm eq}(q,T(t))$. 
Deviations from the equilibrium predictions 
$\alpha = 1$ and $q_{FS} \sim 1/N_z$ can be 
understood analytically applying non-equilibrium mean-field methods such as those developed in Refs. \cite{Lei2019SciAdv,MaireChaix2025arXiv2509.04242}. This  would be a promising direction for understanding non-equilibrium effects in the vortex system, but it goes beyond the aim of this work.

\section*{Conclusion}

By means of three-dimensional simulations of a system of layered interacting elastic lines, we study the impact of finite-thickness and non-equilibrium effects on the hyperuniformity of the solid  phase obtained by slowly cooling from the liquid phase. This study was driven by the aim of improving the interpretation of recent experimental results~\cite{Rumi2019,Besana2024}. 
Concerning the main question of whether the thickness dependence of the hyperuniformity exponent is an equilibrium property or an out-of-equilibrium effect arising from slow dynamics during cooling, our results indicate that non-equilibrium effects play a central role and yield a similar decay of the crossover towards non-hyperuniform fluctuations at large length scales. 
Which mechanism dominates, equilibrium or non-equilibrium finite-thickness effects, depends on the cooling rate, the sample thickness, and the relaxation of long-wavelength modes, the latter being strongly hindered in the presence of disorder.

These findings provide a framework to interpret the finite-thickness crossovers observed in recent works imaging vortex structures in layered superconductors~\cite{Besana2024}. In a broader perspective, this study provides crucial information on how to control the relevant experimental parameters when trying to synthesize hyperuniform structures by cooling in realistic host media, namely having finite size and naturally presenting disorder.


\section*{Appendix}

\begin{figure}[hhh]
        \centering        \includegraphics[width=\linewidth]{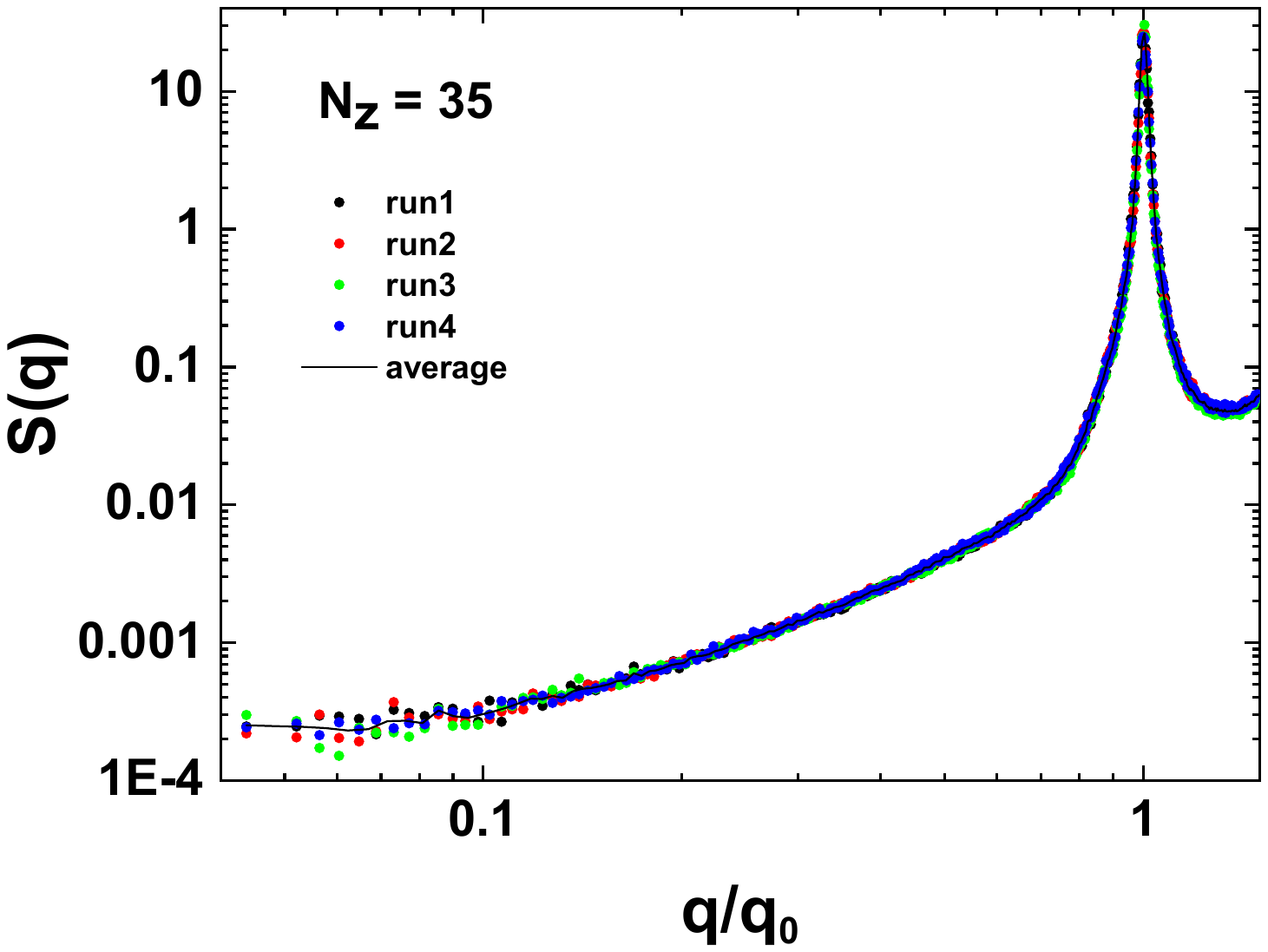}
        \caption{Angularly-averaged structure factor data for $N_z=35$ from 4 independent simulation realizations (points) and average value (black line).}
        \label{fig:Figure11}
    \end{figure}

Figure\,\ref{fig:Figure11} shows $S(q)$ data for  $N_z=35$ from 4 independent simulation realizations (points) and the average of them (blue line). The dispersion in data enhances when decreasing $q$ but nevertheless remains smaller than $10^{-4}$.

\begin{figure}[hhh]
        \centering
        \includegraphics[width=\linewidth]{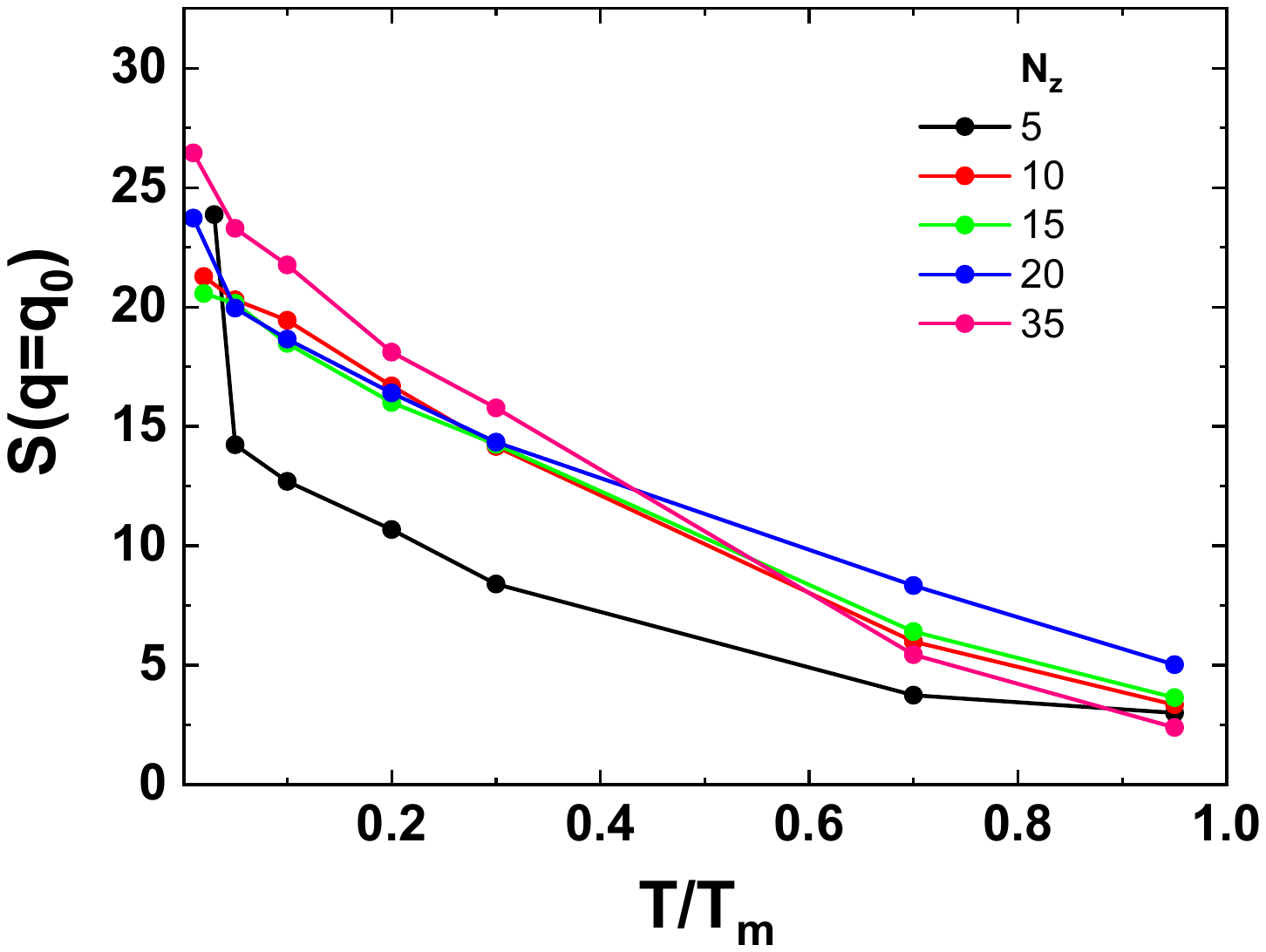}
        \caption{Height of the Bragg peak as a function of reduced temperature $T/T_{\rm m}$ for selected $N_{z}$ values between 5 and 35.}
        \label{fig:Figure12}
    \end{figure}

Figure\,\ref{fig:Figure12} presents data on the height of the peak in the structure factor, $S(q=q_{0})$, as a function of the reduced temperature $T/T_{\rm m}$ for selected $N_z$. Irrespective of $N_{z}$, $S(q=q_{0})$ decreases on increasing the reduced temperature and for $T/T_{\rm m} \sim 1$ the different curves tend to a value in the range 2.5-5. This asymptotic value is roughly the one expected for liquids in view of the Hansen-Verlet criteria stating that the liquid freezes when the first peak of the structure factor reaches a critical value that depends on dimensionality and interactions. In the case of a vortex liquid, Ref.~\onlinecite{Cornaglia1999} shows that this value is around 6.5 and even smaller when going to the 2D limit.

\bibliography{biblio}

\end{document}